\documentclass[notitlepage, twocolumn, letterpaper, prx, longbibliography, 10pt]{revtex4-2}
\usepackage[utf8]{inputenc}
\usepackage{amsmath}
\usepackage{amssymb}
\usepackage{graphicx}
\usepackage{hyperref}
\usepackage{subcaption}
\usepackage[percent]{overpic}
\usepackage{pdfpages}
\usepackage[normalem]{ulem}
\usepackage[switch,columnwise]{lineno}
\usepackage[margin=1 in]{geometry}
\usepackage[object=vectorian]{pgfornament} 

\newcommand{\dd}[1]{\mathrm{d}#1}

\newcommand{\deriv}[2]{\frac{\dd{#1}}{\dd{#2}}}

\DeclareMathOperator{\erf}{Erf}

\makeatletter
\AtBeginDocument{\let\LS@rot\@undefined}
\makeatother

\graphicspath{{images/}}

\begin{document}

\title{Interpretation of generalized Langevin equations}
\author{David Sabin-Miller}
\email{dasami@umich.edu}
\affiliation{Center for the Study of Complex Systems, University of Michigan, Ann Arbor, MI 48109}
\author{Daniel M.~Abrams}
\email{dmabrams@northwestern.edu}
\affiliation{Department of Engineering Sciences and Applied Mathematics, Northwestern University, Evanston, IL, USA}
\affiliation{Department of Physics and Astronomy, Northwestern University, Evanston, IL, USA}

\begin{abstract}
    Many real-world systems exhibit ``noisy'' evolution in time; interpreting their finitely-sampled behavior as arising from continuous-time processes (in the It\^o or Stratonovich sense) has led to significant success in modeling and analysis in a wide variety of fields. Yet such interpretation hinges on a fundamental linear separation of randomness from determinism in the underlying dynamics. Here we propose some theoretical systems which resist easy and self-consistent interpretation into this well-defined class of equations, requiring an expansion of the interpretive framework. We argue that a wider class of stochastic differential equations, where evolution depends nonlinearly on a random or effectively-random quantity, may be consistently interpreted and in fact exhibit finite-time stochastic behavior in line with an equivalent It\^o process, at which point many existing numerical and analytical techniques may be used.
    We put forward a method for this conversion, and demonstrate its use on both a toy system and on a system of direct physical relevance: the velocity of a meso-scale particle suspended in a turbulent fluid. This work enables the theoretical and numerical examination of a wide class of mathematical models which might otherwise be oversimplified due to a lack of appropriate tools.
\end{abstract}

\maketitle

\section{Generalizing Langevin Equations}
Langevin equations are often used to represent theoretical differential behavior for systems exhibiting stochastic dynamics (see, e.g., \cite{coffey2012langevin, gardiner2009stochastic}). These equations have a standard form, which we will aim to generalize:
\begin{equation*}
    \deriv{x}{t} = f(x,t) + g(x,t) \eta_t \, , 
\end{equation*}
where $\eta_t$ represents the ``Gaussian white noise'' term, $\delta$-correlated in continuous time. If $g(x,t)$ exhibits $x$ dependence, such Langevin equations are ill-defined, necessitating a choice of either the It\^o or Stratonovich interpretation---these will differ in the resulting ``drift'' behavior of the system, but both are internally consistent and able to be simulated and analyzed by various techniques \cite{van1981ito, gardiner2009stochastic}. 

Here, we seek to generalize to systems of the form 
\begin{equation} \label{eq:generalsys}
    \deriv{x}{t} = R(x,t), 
\end{equation}
where $R$ is some random variable with some (possibly non-Gaussian) probability distribution over the domain. We argue that, with the proper conversion procedure based on the central limit theorem \cite{fischer2010history}, these Langevin-type systems may be reduced to equivalent It\^o behavior, allowing for consistent simulation and theoretical analysis. 

As a motivating example, we start by highlighting the difference between two similar-looking Langevin-type equations:
 \begin{align}
    \deriv{x}{t} &= -x^3 + \eta_t, \label{eq:langevin}\\
    \deriv{x}{t} &= -(x + \eta_t)^3.  \label{eq:nonlin_langevin}
\end{align}

Equation \eqref{eq:langevin} is a classic Langevin equation with cubic attraction towards zero and diffusive noise---easily interpreted (in either the It\^o or Stratonovich sense) as the stochastic differential equation (SDE) $ \dd{x} = -x^3 \dd{t} + \dd{W}$ (where $\dd{W}$ represents the usual derivative of a Wiener process), enabling all the analytical and numerical options that entails.

Equation \eqref{eq:nonlin_langevin}, however, is notably different in that the nonlinear cubing operation happens to a fundamentally random quantity, linking the deterministic and random parts of the equation. Na\"ive numerical simulation simply converges to deterministic behavior as the time-step shrinks, since the fluctuations average out before $x$ changes considerably. If timestep-independent stochastic behavior is desired, we must develop a new consistent and coherent interpretation of this equation. 

We note that the notation of Eq,~\eqref{eq:nonlin_langevin} would be better written as
 \begin{align}
    \deriv{x}{t} &= -X^3, \textrm{ where } X \sim N(x,1). \label{eq:nonlin_langevin_X}
\end{align}
This is because the ``randomness'' term $\eta$ requires an explicit distribution in this case, rather than leaning on the Central Limit Theorem to abstract that information away
. While Eq.~\eqref{eq:langevin} would behave identically regardless of which distribution $\eta$ represents, as long as it has zero mean and standard deviation 1---a broad equivalence class which allows the use of the small-d$t$ limit, the normal distribution, without loss of generality---the nonlinear operation being applied in \eqref{eq:nonlin_langevin} requires an explicit choice of this underlying ``noise,'' which we might, for example, pick to be a normal distribution which is then distorted by cubing as indicated in Eq.~\eqref{eq:nonlin_langevin_X}.
 
 \subsection{Potential Applications}
 ``Baked-in'' stochasticity of this type might arise in a variety of physical modeling scenarios. For example, nonlinear drag forces acting on a macroscopic object in a turbulent flow would cause velocity to evolve according to this type of Langevin equation, with ``noise'' coming from rapidly fluctuating relative fluid velocity---including, e.g., viscous drag on a cylinder in a turbulent wake \cite{Hover_2001}. We compute results for this velocity distribution, and its stark difference from a na\"ive approach, at the end of this section. Physical systems with nonlinear feedback based on rapidly fluctuating quantities or quantities subject to random measurement error would also be of this type. Inasmuch as measurement error acts as independent random variation of a quantity, the behavior of simulated or artificially-forced feedback-based dynamical systems would also benefit from this analysis. Our interest was motivated by an earlier model for individuals reacting to a stochastic political environment \cite{DSM2020}. A variety of other physics-inspired  nonlinear models of complex real-world phenomena may also share this form. 

We note that the systems we are concerned with differ from other ways in which nonlinearity can arise in stochastic systems, for example in the deterministic part (e.g.~\cite{Kappen2005control}) or when $x$-dependence appears multiplied by the stochastic quantity (e.g.,~\cite{zhang2005robust, berman2006control}), or when functions are applied to a continuous random-\textit{walking} quantity (as It\^o's lemma would handle~\cite{gardiner2009stochastic}) rather than the uncertain/noisy quantity itself. Certain specific problems exhibiting nonlinear dependence on stochastic quantities have been examined \cite{kraichnan1961dynamics}, but a general theory of this class of stochastic equations has not been developed.

\subsection{The Proposed Equivalence}
We seek to bridge the gap from the theoretical, possibly non-Gaussian ``intrinsic" noise (represented by the distribution $R$ in Eq.~\eqref{eq:generalsys}) to some equivalent emergent system which is well-defined, self-consistent, and able to be simulated. 

Our argument is based on the consideration that over any finite time-scale, a theoretical system such as Eq.~\eqref{eq:nonlin_langevin} will have experienced a large enough number of nearly-independent increments that the Generalized Central Limit Theorem should apply \cite{kolmogorov1954limitDistributions}. That is, the net increment over any finite time must be drawn from the family of \textit{stable distributions}, or---if the ``intrinsic" noise represented by the differential update distribution itself has finite variance---a Gaussian distribution in particular \cite{kolmogorov1954limitDistributions}. This intuitively dovetails with the more practically-motivated necessary condition that, in the numerical simulation of any continuous-time system, its behavior must not depend sensitively on the simulated timestep; that is, one relatively large step must result in the same distribution (in an ensemble average sense) as the commensurate number of arbitrarily small steps. 

We proceed henceforth with the assumption of finite underlying variance. This means that the net increment over any small but finite time must be drawn from a Gaussian distribution with mean equal to the mean of the underlying process. We may also choose this Gaussian's distribution's variance per unit time to likewise match the underlying process, maintaining consistency with the classic Langevin-It\^o conversion and agreement in standard cases. 

By this reasoning, we argue that every such stochastic process with finite variance is in fact \textit{equivalent} to an It\^o SDE over any finite time-scale: in particular, the SDE with deterministic part matching the ``true'' distribution's mean behavior and random part matching its standard deviation. We note that this is not a one-to-one mapping, but rather many-to-one: any stochastic process with the same mean and standard deviation would behave identically, and thus be represented by the same It\^o SDE.  

That is, for a general stochastic system of the form
\begin{equation*}
    \frac{\dd{x}}{\dd{t}} = R(x,t) \sim P(r|x,t),
\end{equation*}
where $R$ is some finite-variance stochastic quantity dependent on $x$ and $\delta$-correlated in time, with distribution $P$, one should simulate the It\^o SDE
\begin{align*}
& \qquad  \dd{x} = F(x,t) \dd{t} + G(x,t) \dd{W}, \quad \textrm{where} \\
    &F(x,t) = \textrm{mean}\left[R(x,t)\right] = \int\limits_{-\infty}^{\infty} r P(r|x,t) \dd{r}, \\
    &G(x,t) = \textrm{std}\left[R(x,t)\right] = \sqrt{ \int\limits_{-\infty}^{\infty} \left[r-F(x,t)\right]^2 P(r|x,t) \dd{r} },
\end{align*}
if these quantities exist. We will limit ourselves to stationary and autonomous processes (i.e., $F(x,t)=F(x)$ and $G(x,t)=G(x)$) for demonstration from this point forward, but the theory should extend to non-stationary processes.

Once we have this It\^o equation, we may use standard numerical integration techniques for individual trajectories, or convert the system to a Fokker-Planck form and evolve the solution's probability distribution $\rho(x)$ directly, with
\begin{equation*}
    \frac{\partial \rho (x,t)}{\partial t} = -\frac{\partial}{\partial x} \left[ F(x) \rho(x,t) \right] + \frac{1}{2} \frac{\partial^2}{\partial x^2} \left[ G(x)^2 \rho(x,t) \right]\;.
\end{equation*}

As an example, we will now examine a slightly generalized version of Eq.~\eqref{eq:nonlin_langevin} to determine the effect of noise with arbitrary constant amplitude $\sigma$:
\begin{align}
    \deriv{x}{t} &= -X^3, \textrm{ where } X \sim N(x,\sigma). \label{eq:nonlin_new_sigma}
\end{align}
In section S1 of the Supplemental Material (SM), we examine a yet more general version of this attractor with arbitrary positive-integer exponent, but for illustration and concreteness henceforth focus on this cubic nonlinear-stochastic attractor.  Using the shorthand notation $N(r|\mu,\sigma) = e^{\frac{-(r-\mu)^2}{2\sigma^2}} / (\sigma\sqrt{2 \pi})$, we have:
\begin{align*}
    F(x|\sigma) &= \int\limits_{-\infty}^{\infty} -r^3 N(r|x,\sigma) \dd{r}\\
    &= -x^3 - 3 \sigma^2 x
\end{align*}
and 
\begin{align*}
    G(x|\sigma) &= \sqrt{\int\limits_{-\infty}^{\infty} [-r^3-F(x)]^2 N(r|x,\sigma) \dd{r}}\\
    &= \sqrt{9 \sigma^2 x^4 + 36 \sigma^4 x^2 +15\sigma^6  }.
\end{align*}
%
%
So we argue that the system 
\begin{align*}
    \frac{\dd{x}}{\dd{t}} = -X^3, \textrm{ where } X \sim N(x,\sigma)
\end{align*}
is equivalent to the It\^o SDE 
\begin{align}
    \dd{x} &= (-x^3-3\sigma^2 x)\ \dd{t}  \nonumber \\
    & \qquad + \sqrt{15\sigma^6 + 36 \sigma^4 x^2 +9 \sigma^2 x^4}\ \dd{W}, \label{eq:nIto_form_of_cubic}
\end{align}
which is amenable to various methods of simulation and analysis like any other It\^o equation. We note that this It\^o equation is significantly different from anything one might obtain from the similar-looking but simply additive true-Langevin form in Eq.~\eqref{eq:langevin}.

To reiterate: a na\"ive interpretation of Eq.~\eqref{eq:nonlin_new_sigma} (simply ``expanding'' $-(x+\sigma \eta_t)^3$ and taking the pure-deterministic and pure-stochastic terms) might lead to the It\^o SDE 
\begin{equation}
    \dd{x} = -x^3 \dd{t} + \sigma^3 \dd{W}\;,
    \label{eq:naivecubic}
\end{equation}
which has completely different physical behavior than our proposed interpretation in Eq.~\eqref{eq:nIto_form_of_cubic}\footnote{In our ``na\"ive'' interpretation, we assume expansion of $(x+\sigma \eta)^3$ would be approximated with constant noise of amplitude $\sigma^3$, though the amplitude of constant noise could also be taken to be a fitted constant.}.  Basic properties like the variance of the equilibrium distribution differ, with divergence possible in Eq.~\eqref{eq:nIto_form_of_cubic} but not in Eq.~\eqref{eq:naivecubic}.  This has significant implications for all types of stochastic models used throughout physics.

\subsection{Application to Drag in Turbulent Fluid}
As an illustrative physical example, we consider the regime of quadratic drag with rapidly varying relative fluid velocity---of relevance to the behavior of particles in well-developed turbulence. In the one-dimensional case without stochasticity, relative velocity $v$ would evolve according to 
$$\textrm{d}v/\textrm{d}t = -c v|v|$$
(here the constant $c$ sets the time scale, and we set it to $1$ henceforth).  When rapid random relative-velocity fluctuations are included, we have:
\begin{align}
    \deriv{v}{t} &= -V|V| \ , \textrm{ \quad where } V \sim N(v,\sigma). \label{eq:drag_langevin}
\end{align}
This might na\"ively be modeled by the It\^o equation 
\begin{equation}
    \dd{v} = -v|v| \dd{t} + \sigma^2 \dd{W}\;,
    \label{eq:drag_naive_Ito}
\end{equation}
which has an exact solution for its steady-state probability distribution
\begin{equation}
    p(v) = \frac{C}{\sigma^{2/3}} \exp{\left(\frac{-2 |v^3|}{3 \sigma^2}\right)}\;,
    \label{eq:drag_naive_exact_sol}
\end{equation}
where $C$ is a normalization constant, namely $3^{7/6}\Gamma(2/3)/(2^{5/3} \pi)$.

\begin{figure}[t!] 
\centering
	\includegraphics[width=\columnwidth]{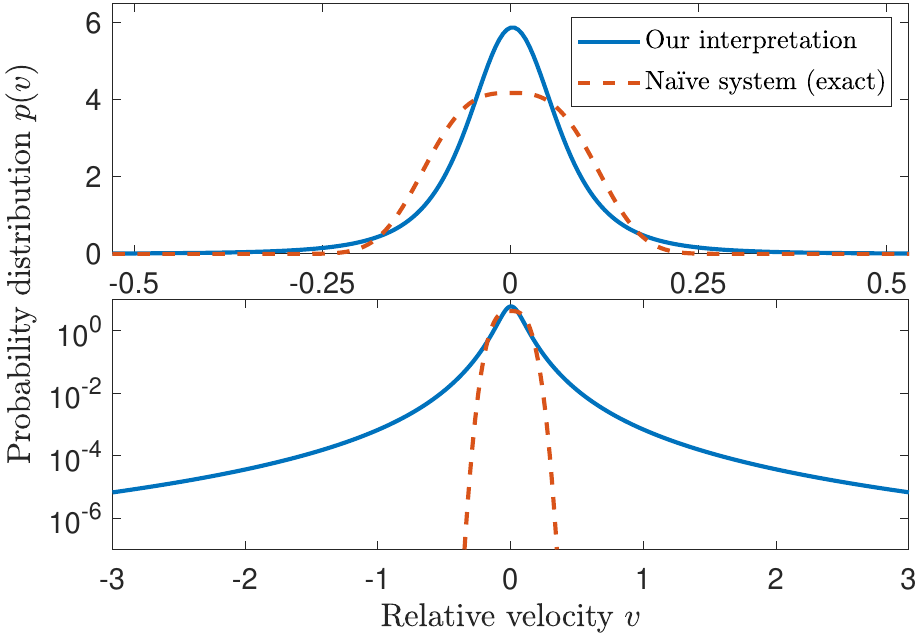}
	\caption{\textbf{Equilibrium velocity distributions.} Comparison of equilibrium distributions for the drag system in Eq.~\eqref{eq:drag_langevin} with $\sigma = 0.2$, computed by Fokker-Planck integration of our proposed behavior \eqref{eq:drag_better_Ito} and compared to the exact solution \eqref{eq:drag_naive_exact_sol} for a na\"ive interpretation of the system's behavior.\textbf{Top:} Linear scale. \textbf{Bottom:} Log-scale view, with wider x domain to show clear differences in implied distributions.
	} 
	\label{fig:drag_diff}
\end{figure}

But we propose that this system is more faithfully modeled by using our proposed conversion, which yields
\begin{equation}
    \dd{v} = F_2(v|\sigma) \dd{t} + G_2(v|\sigma) \dd{W}\;, 
    \label{eq:drag_better_Ito}
\end{equation}
where
\begin{align*}
    F_2(v|\sigma) &= -\left(\sigma^2 + v^2\right)\erf\left(\frac{v}{\sigma \sqrt{2}}\right) 
    - \sqrt{\frac{2}{\pi}} x \sigma e^{\frac{-v^2}{2 \sigma^2}} \;,\\
 G_2(v|\sigma) &= \sqrt{v^4 + 6v^2 \sigma^2 +3 \sigma^4 + 3[F_2(v|\sigma)]^2 }
\end{align*}
(computation details in section S1 of the SM). The significant difference in behavior between these systems is illustrated in
in Fig.~\ref{fig:drag_diff}. 

\section{Discussion and Limitations}
The first proposition of this paper---the argument for Ito-equivalency of nonlinear Langevin-type systems---is really a proposed definition rather than a theoretical result.  Like Langevin equations themselves, the notation is simple and intuitive, but solid mathematical interpretation requires the use of the more rigorous notation, and we propose that interpretation in terms of It\^o calculus---though this could easily be reinterpreted in the Stratonovich sense as convention or data favor. 

We apply logic based on the central limit theorem for finite-variance random variables, but the Langevin noise terms are not regular random variables and their variance may not be well-defined or finite. If variance is treated as well-defined but not finite, other (non-Gaussian) stable distributions per time-step may arise, rather than normally distributed It\^o time-steps. 

We also note the perhaps-undesirable sensitivity to the assumption of Gaussian underlying noise in Eq.~\eqref{eq:generalsys}.  
In particular, the assumption that $\eta_t$ is normally distributed may be incorrect for some systems with biased or irregularly shaped noise, and if the noise shape is known it should be used.

\section{Conclusions}
We have shown that a class of ``nonlinear-stochastic'' Langevin equations may be interpreted such that they have well-defined behavior after conversion to an equivalent It\^o system. This type of equilibrium may be more general than initially apparent, since nearly any isolated attracting fixed point is locally well-approximated by equations of this form.

We have applied this theory to cubic and quadratic nonlinear attracting fixed points and arrived at appropriate equivalent It\^o systems. These systems notably differ from any na\"ive system with $x$-independent noise, and Fokker-Planck numerical integration confirms that the implied equilibria of the quadratic case (a model of the velocity of a particle experiencing drag in a turbulent fluid) are quite different. In other work \cite{DSM2025equilibrium} we analyze the equilibria of the cubic system via moment relations and find that its higher moments diverge above a critical value of $\sigma$, and show that simulations bear out this analysis.

This conversion technique should lead to more faithful physical modeling, yielding qualitatively different behavior when compared to simplifications which transform a deterministic quantity and add noise afterward. 
Conversely, our reasoning also leads to the implication that apparent, observable It\^o behavior might be emergent from any number of nonlinear Langevin processes at the theoretical level.

\begin{center}
  \pgfornament[width = 0.8\columnwidth, color = black]{88}
\end{center}

\begin{acknowledgments}
The authors thank Bill Kath for useful conversation, Gary Nave for help with relevant literature, and the National Science Foundation for support through the Graduate Research Fellowship Program.
\end{acknowledgments}

%


\begin{thebibliography}{20}%
\makeatletter
\providecommand \@ifxundefined [1]{%
 \@ifx{#1\undefined}
}%
\providecommand \@ifnum [1]{%
 \ifnum #1\expandafter \@firstoftwo
 \else \expandafter \@secondoftwo
 \fi
}%
\providecommand \@ifx [1]{%
 \ifx #1\expandafter \@firstoftwo
 \else \expandafter \@secondoftwo
 \fi
}%
\providecommand \natexlab [1]{#1}%
\providecommand \enquote  [1]{``#1''}%
\providecommand \bibnamefont  [1]{#1}%
\providecommand \bibfnamefont [1]{#1}%
\providecommand \citenamefont [1]{#1}%
\providecommand \href@noop [0]{\@secondoftwo}%
\providecommand \href [0]{\begingroup \@sanitize@url \@href}%
\providecommand \@href[1]{\@@startlink{#1}\@@href}%
\providecommand \@@href[1]{\endgroup#1\@@endlink}%
\providecommand \@sanitize@url [0]{\catcode `\\12\catcode `\$12\catcode
  `\&12\catcode `\#12\catcode `\^12\catcode `\_12\catcode `\%12\relax}%
\providecommand \@@startlink[1]{}%
\providecommand \@@endlink[0]{}%
\providecommand \url  [0]{\begingroup\@sanitize@url \@url }%
\providecommand \@url [1]{\endgroup\@href {#1}{\urlprefix }}%
\providecommand \urlprefix  [0]{URL }%
\providecommand \Eprint [0]{\href }%
\providecommand \doibase [0]{https://doi.org/}%
\providecommand \selectlanguage [0]{\@gobble}%
\providecommand \bibinfo  [0]{\@secondoftwo}%
\providecommand \bibfield  [0]{\@secondoftwo}%
\providecommand \translation [1]{[#1]}%
\providecommand \BibitemOpen [0]{}%
\providecommand \bibitemStop [0]{}%
\providecommand \bibitemNoStop [0]{.\EOS\space}%
\providecommand \EOS [0]{\spacefactor3000\relax}%
\providecommand \BibitemShut  [1]{\csname bibitem#1\endcsname}%
\let\auto@bib@innerbib\@empty
\bibitem [{\citenamefont {Coffey}\ and\ \citenamefont
  {Kalmykov}(2012)}]{coffey2012langevin}%
  \BibitemOpen
  \bibfield  {author} {\bibinfo {author} {\bibfnamefont {W.}~\bibnamefont
  {Coffey}}\ and\ \bibinfo {author} {\bibfnamefont {Y.~P.}\ \bibnamefont
  {Kalmykov}},\ }\href@noop {} {\emph {\bibinfo {title} {The {L}angevin
  equation: {W}ith applications to stochastic problems in physics, chemistry
  and electrical engineering}}},\ Vol.~\bibinfo {volume} {27}\ (\bibinfo
  {publisher} {World Scientific},\ \bibinfo {year} {2012})\BibitemShut
  {NoStop}%
\bibitem [{\citenamefont {Gardiner}(2009)}]{gardiner2009stochastic}%
  \BibitemOpen
  \bibfield  {author} {\bibinfo {author} {\bibfnamefont {C.}~\bibnamefont
  {Gardiner}},\ }\href@noop {} {\emph {\bibinfo {title} {Stochastic
  {M}ethods}}},\ Vol.~\bibinfo {volume} {4}\ (\bibinfo  {publisher} {Springer
  Berlin},\ \bibinfo {year} {2009})\BibitemShut {NoStop}%
\bibitem [{\citenamefont {Van~Kampen}(1981)}]{van1981ito}%
  \BibitemOpen
  \bibfield  {author} {\bibinfo {author} {\bibfnamefont {N.~G.}\ \bibnamefont
  {Van~Kampen}},\ }\bibfield  {title} {\bibinfo {title} {It{\^o} versus
  {S}tratonovich},\ }\href@noop {} {\bibfield  {journal} {\bibinfo  {journal}
  {Journal of Statistical Physics}\ }\textbf {\bibinfo {volume} {24}},\
  \bibinfo {pages} {175} (\bibinfo {year} {1981})}\BibitemShut {NoStop}%
\bibitem [{\citenamefont {Fischer}(2010)}]{fischer2010history}%
  \BibitemOpen
  \bibfield  {author} {\bibinfo {author} {\bibfnamefont {H.}~\bibnamefont
  {Fischer}},\ }\href@noop {} {\emph {\bibinfo {title} {A history of the
  central limit theorem: {F}rom classical to modern probability theory}}}\
  (\bibinfo  {publisher} {Springer Science \& Business Media},\ \bibinfo {year}
  {2010})\BibitemShut {NoStop}%
\bibitem [{\citenamefont {Hover}\ and\ \citenamefont
  {Triantafyllou}(2001)}]{Hover_2001}%
  \BibitemOpen
  \bibfield  {author} {\bibinfo {author} {\bibfnamefont {F.}~\bibnamefont
  {Hover}}\ and\ \bibinfo {author} {\bibfnamefont {M.}~\bibnamefont
  {Triantafyllou}},\ }\bibfield  {title} {\bibinfo {title} {Galloping response
  of a cylinder with upstream wake interference},\ }\href
  {https://doi.org/https://doi.org/10.1006/jfls.2000.0364} {\bibfield
  {journal} {\bibinfo  {journal} {Journal of Fluids and Structures}\ }\textbf
  {\bibinfo {volume} {15}},\ \bibinfo {pages} {503 } (\bibinfo {year}
  {2001})}\BibitemShut {NoStop}%
\bibitem [{\citenamefont {Sabin-Miller}\ and\ \citenamefont
  {Abrams}(2020)}]{DSM2020}%
  \BibitemOpen
  \bibfield  {author} {\bibinfo {author} {\bibfnamefont {D.}~\bibnamefont
  {Sabin-Miller}}\ and\ \bibinfo {author} {\bibfnamefont {D.~M.}\ \bibnamefont
  {Abrams}},\ }\bibfield  {title} {\bibinfo {title} {When pull turns to shove:
  A continuous-time model for opinion dynamics},\ }\href
  {https://doi.org/10.1103/PhysRevResearch.2.043001} {\bibfield  {journal}
  {\bibinfo  {journal} {Physical Review Research}\ }\textbf {\bibinfo {volume}
  {2}},\ \bibinfo {pages} {043001} (\bibinfo {year} {2020})}\BibitemShut
  {NoStop}%
\bibitem [{\citenamefont {Kappen}(2005)}]{Kappen2005control}%
  \BibitemOpen
  \bibfield  {author} {\bibinfo {author} {\bibfnamefont {H.~J.}\ \bibnamefont
  {Kappen}},\ }\bibfield  {title} {\bibinfo {title} {Linear theory for control
  of nonlinear stochastic systems},\ }\href
  {https://doi.org/10.1103/PhysRevLett.95.200201} {\bibfield  {journal}
  {\bibinfo  {journal} {Physical Review Letters}\ }\textbf {\bibinfo {volume}
  {95}},\ \bibinfo {pages} {200201} (\bibinfo {year} {2005})}\BibitemShut
  {NoStop}%
\bibitem [{\citenamefont {Zhang}\ \emph {et~al.}(2005)\citenamefont {Zhang},
  \citenamefont {Chen},\ and\ \citenamefont {Tseng}}]{zhang2005robust}%
  \BibitemOpen
  \bibfield  {author} {\bibinfo {author} {\bibfnamefont {W.}~\bibnamefont
  {Zhang}}, \bibinfo {author} {\bibfnamefont {B.-S.}\ \bibnamefont {Chen}},\
  and\ \bibinfo {author} {\bibfnamefont {C.-S.}\ \bibnamefont {Tseng}},\
  }\bibfield  {title} {\bibinfo {title} {Robust {H}-infinity filtering for
  nonlinear stochastic systems},\ }\href@noop {} {\bibfield  {journal}
  {\bibinfo  {journal} {IEEE Transactions on Signal Processing}\ }\textbf
  {\bibinfo {volume} {53}},\ \bibinfo {pages} {589} (\bibinfo {year}
  {2005})}\BibitemShut {NoStop}%
\bibitem [{\citenamefont {Berman}\ and\ \citenamefont
  {Shaked}(2006)}]{berman2006control}%
  \BibitemOpen
  \bibfield  {author} {\bibinfo {author} {\bibfnamefont {N.}~\bibnamefont
  {Berman}}\ and\ \bibinfo {author} {\bibfnamefont {U.}~\bibnamefont
  {Shaked}},\ }\bibfield  {title} {\bibinfo {title} {{H}-infinity-like control
  for nonlinear stochastic systems},\ }\href@noop {} {\bibfield  {journal}
  {\bibinfo  {journal} {Systems \& Control Letters}\ }\textbf {\bibinfo
  {volume} {55}},\ \bibinfo {pages} {247} (\bibinfo {year} {2006})}\BibitemShut
  {NoStop}%
\bibitem [{\citenamefont {Kraichnan}(1961)}]{kraichnan1961dynamics}%
  \BibitemOpen
  \bibfield  {author} {\bibinfo {author} {\bibfnamefont {R.~H.}\ \bibnamefont
  {Kraichnan}},\ }\bibfield  {title} {\bibinfo {title} {Dynamics of nonlinear
  stochastic systems},\ }\href@noop {} {\bibfield  {journal} {\bibinfo
  {journal} {Journal of Mathematical Physics}\ }\textbf {\bibinfo {volume}
  {2}},\ \bibinfo {pages} {124} (\bibinfo {year} {1961})}\BibitemShut {NoStop}%
\bibitem [{\citenamefont {Gnedenko}\ and\ \citenamefont
  {Kolmogorov}(1954)}]{kolmogorov1954limitDistributions}%
  \BibitemOpen
  \bibfield  {author} {\bibinfo {author} {\bibfnamefont {B.~V.}\ \bibnamefont
  {Gnedenko}}\ and\ \bibinfo {author} {\bibfnamefont {A.~N.}\ \bibnamefont
  {Kolmogorov}},\ }\href@noop {} {\emph {\bibinfo {title} {Limit distributions
  for sums of independent random variables}}},\ Addison-Wesley Mathematics
  Series\ (\bibinfo  {publisher} {Addison-Wesley},\ \bibinfo {address}
  {Cambridge, MA},\ \bibinfo {year} {1954})\ pp.\ \bibinfo {pages} {ix+264},\
  \bibinfo {note} {translated and annotated by K. L. Chung. With an Appendix by
  J. L. Doob.}\BibitemShut {Stop}%
\bibitem [{Note1()}]{Note1}%
  \BibitemOpen
  \bibinfo {note} {In our ``naive'' interpretation, we assume expansion of
  $(x+\sigma \eta )^3$ would be approximated with constant noise of amplitude
  $\sigma ^3$, though the amplitude of constant noise could also be taken to be
  a fitted constant.}\BibitemShut {Stop}%
\bibitem [{\citenamefont {Maruyama}(1955)}]{maruyama1955continuous}%
  \BibitemOpen
  \bibfield  {author} {\bibinfo {author} {\bibfnamefont {G.}~\bibnamefont
  {Maruyama}},\ }\bibfield  {title} {\bibinfo {title} {Continuous {M}arkov
  processes and stochastic equations},\ }\href@noop {} {\bibfield  {journal}
  {\bibinfo  {journal} {Rendiconti del Circolo Matematico di Palermo}\ }\textbf
  {\bibinfo {volume} {4}},\ \bibinfo {pages} {48} (\bibinfo {year}
  {1955})}\BibitemShut {NoStop}%
\bibitem [{Note2()}]{Note2}%
  \BibitemOpen
  \bibinfo {note} {Changes in the order of integration will always be allowable
  for finite $\mu _2$.}\BibitemShut {Stop}%
\bibitem [{Note3()}]{Note3}%
  \BibitemOpen
  \bibinfo {note} {We note that the earlier assumption of finite underlying
  stochastic-process variance was only necessary to arrive at the It\^o SDE
  from the initial Langevin-type equation; our analysis of the It\^o SDE itself
  does not rely on that assumption.}\BibitemShut {Stop}%
\bibitem [{Note4()}]{Note4}%
  \BibitemOpen
  \bibinfo {note} {For symmetric equilibria like our cubic example, odd moments
  are all zero.}\BibitemShut {Stop}%
\bibitem [{\citenamefont {Uhlenbeck}\ and\ \citenamefont
  {Ornstein}(1930)}]{uhlenbeck1930theory}%
  \BibitemOpen
  \bibfield  {author} {\bibinfo {author} {\bibfnamefont {G.~E.}\ \bibnamefont
  {Uhlenbeck}}\ and\ \bibinfo {author} {\bibfnamefont {L.~S.}\ \bibnamefont
  {Ornstein}},\ }\bibfield  {title} {\bibinfo {title} {On the theory of the
  {B}rownian motion},\ }\href@noop {} {\bibfield  {journal} {\bibinfo
  {journal} {Physical Review}\ }\textbf {\bibinfo {volume} {36}},\ \bibinfo
  {pages} {823} (\bibinfo {year} {1930})}\BibitemShut {NoStop}%
\bibitem [{\citenamefont {Wang}\ and\ \citenamefont
  {Uhlenbeck}(1945)}]{wang1945theory}%
  \BibitemOpen
  \bibfield  {author} {\bibinfo {author} {\bibfnamefont {M.~C.}\ \bibnamefont
  {Wang}}\ and\ \bibinfo {author} {\bibfnamefont {G.~E.}\ \bibnamefont
  {Uhlenbeck}},\ }\bibfield  {title} {\bibinfo {title} {On the theory of the
  {B}rownian motion {II}},\ }\href@noop {} {\bibfield  {journal} {\bibinfo
  {journal} {Reviews of Modern Physics}\ }\textbf {\bibinfo {volume} {17}},\
  \bibinfo {pages} {323} (\bibinfo {year} {1945})}\BibitemShut {NoStop}%
\bibitem [{\citenamefont {Van~Kampen}(2007)}]{van1992stochastic}%
  \BibitemOpen
  \bibfield  {author} {\bibinfo {author} {\bibfnamefont {N.~G.}\ \bibnamefont
  {Van~Kampen}},\ }\href@noop {} {\emph {\bibinfo {title} {Stochastic processes
  in physics and chemistry}}}\ (\bibinfo  {publisher} {Elsevier},\ \bibinfo
  {year} {2007})\BibitemShut {NoStop}%
\bibitem [{\citenamefont {Gillespie}(1991)}]{gillespie1991markov}%
  \BibitemOpen
  \bibfield  {author} {\bibinfo {author} {\bibfnamefont {D.~T.}\ \bibnamefont
  {Gillespie}},\ }\href@noop {} {\emph {\bibinfo {title} {Markov processes: an
  introduction for physical scientists}}}\ (\bibinfo  {publisher} {Elsevier},\
  \bibinfo {year} {1991})\BibitemShut {NoStop}%
\bibitem [{\citenamefont {Sabin-Miller}\ and\ \citenamefont {Abrams}(2025)}]{DSM2025equilibrium}%
  \BibitemOpen
  \bibfield  {author} {\bibinfo {author} {\bibfnamefont {David}\ \bibnamefont {Sabin-Miller}}\ and\ \bibinfo {author} {\bibfnamefont {Daniel~M.}\ \bibnamefont {Abrams}},\ }\href {https://arxiv.org/abs/2502.00918} {\enquote {\bibinfo {title} {Equilibrium moment analysis of {I}t\^o {SDE}s},}\ } (\bibinfo {year} {2025}),\ \Eprint {http://arxiv.org/abs/2502.00918} {arXiv:2502.00918 [math.DS]} \BibitemShut {NoStop}%
\end{thebibliography}

\pagebreak 
\newpage 
\clearpage



\section*{Supplementary material}

# Supplemental Material: Modeling and analysis of systems with nonlinear functional dependence on random quantities

David Sabin-Miller*

*Department of Engineering Sciences and Applied Mathematics,
Northwestern University, Evanston, IL, USA*

Daniel M. Abrams

*Department of Engineering Sciences and Applied Mathematics,
Northwestern University, Evanston, IL, USA
Northwestern Institute for Complex Systems,
Northwestern University, Evanston, IL, USA and
Department of Physics and Astronomy, Northwestern University, Evanston, IL, USA*

## S1. GENERALIZATION TO POSITIVE-INTEGER ATTRACTORS

Here we examine a generalization of the nonlinear attracting system from the main text: the $n^{\text{th}}$-order attracting fixed point. As in the main text, the nonlinear attracting function is applied to a Gaussian random variable (in the Langevin sense) centered on the current value (using $N(r; x, \sigma)$ as shorthand for the Gaussian pdf with mean $x$ and standard deviation $\sigma$):

$$r := x + \sigma\eta, \text{ i.e. } r \sim N(r|x, \sigma)$$

$$\frac{\mathrm{d}x}{\mathrm{d}t} = -\operatorname{sgn}(r)|r|^n, \ n \in \mathbf{Z}^+ . \tag{1}$$

Using the proposed Itô conversion from the main text, this should be equivalent to the system

$$\mathrm{d}x = F(x|\sigma, n)\mathrm{d}t + G(x|\sigma, n)\mathrm{d}W, \text{ where} \tag{2}$$

$$F(x|\sigma, n) = \left\langle \frac{\mathrm{d}x}{\mathrm{d}t} \right\rangle = \int_{-\infty}^{\infty} \left[ -\operatorname{sgn}(r)|r|^n \right] N(r|x, \sigma)\mathrm{d}r \tag{3}$$

$$G(x|\sigma, n) = \operatorname{std}\left( \frac{\mathrm{d}x}{\mathrm{d}t} \right) = \sqrt{\int_{-\infty}^{\infty} [-\operatorname{sgn}(r)|r|^n - F(x|\sigma, n)]^2 N(r|x, \sigma)\mathrm{d}r}. \tag{4}$$

We first note that $G$ can be easily computed in terms of $F$:

$$G(x|\sigma, n) = \sqrt{\int_{-\infty}^{\infty} \left[ -\operatorname{sgn}(r)|r|^n - F(x|\sigma, n) \right]^2 N(r|x, \sigma)\mathrm{d}r}$$

$$= \sqrt{\int_{-\infty}^{\infty} \left[ r^{2n} + 2\operatorname{sgn}(r)|r|^n F(x|\sigma, n) + F^2(x|\sigma, n) \right] N(r|x, \sigma)\mathrm{d}r}$$

$$= \sqrt{\int_{-\infty}^{\infty} r^{2n} N(r|x, \sigma)\mathrm{d}r + 2F(x|\sigma, n) \int_{-\infty}^{\infty} \operatorname{sgn}(r)|r|^n N(r|x, \sigma) + F^2(x|\sigma, n) \int_{-\infty}^{\infty} N(r|x, \sigma)\mathrm{d}r}$$

$$= \sqrt{\mu_{2n,x,\sigma} + 3F^2(x|\sigma, n)},$$

* davidsabinmiller@u.northwestern.edu

where the first term is not quite $F(x|\sigma, 2n)$—due to the sign difference—but rather the (much simpler) $2n^{\text{th}}$ non-central moment of the normal distribution $N(r|x, \sigma)$:

$$\mu_{2n,x,\sigma} := \int_{-\infty}^{\infty} r^{2n} N(r|x, \sigma) \mathrm{d}r$$

$$= \int_{-\infty}^{\infty} (x+z)^{2n} N(z|0, \sigma) \mathrm{d}z$$

$$= \sum_{i=0}^{2n} \binom{2n}{i} x^{2n-i} \int_{-\infty}^{\infty} z^i N(z|0, \sigma) \mathrm{d}z$$

$$= \sum_{j=0}^{n} \binom{2n}{2j} (2j-1)!! x^{2n-2j} \sigma^{2j} \ .$$

We now seek $F$. If $n$ is odd, this calculation is simply the non-central moment again:

$$F(x|\sigma, n) = \int_{-\infty}^{\infty} -r^n \left[ \frac{1}{\sigma\sqrt{2\pi}} e^{\frac{-(r-x)^2}{2\sigma^2}} \right] \mathrm{d}r \ \text{ if } n \text{ odd}$$

$$= \sum_{\substack{i=1 \\ i \text{ odd}}}^{n} \binom{n}{i} (i-1)!! x^{n-i} \sigma^i \tag{5}$$

$$= \sum_{\substack{i=1 \\ i \text{ odd}}}^{n} A_{n,i} \ x^{n-i} \sigma^i, \tag{6}$$

where $A_{n,i} := \binom{n}{i}(i-1)!! = \dfrac{n!}{i!!(n-i)!} \ .$ \tag{7}

However if $n$ is even, we must split the integral and the boundary terms no longer cancel due to the sign difference:

$$F(x|\sigma, n) = \int_{-\infty}^{\infty} \left[ -\operatorname{sgn}(r)|r|^n \right] N(r; x, \sigma) \mathrm{d}r$$

$$= \int_{-\infty}^{0} r^n N(r; x, \sigma) \mathrm{d}r - \int_{0}^{\infty} r^n N(r; x, \sigma) \mathrm{d}r$$

$$= \int_{-\infty}^{\infty} r^n N(r; x, \sigma) \mathrm{d}r - 2\int_{0}^{\infty} r^n N(r; x, \sigma) \mathrm{d}r$$

$$= \sum_{\substack{i=0 \\ i \text{ even}}}^{n} A_{n,i} \ x^{n-i} \sigma^i - 2\int_{-x}^{\infty} (x+w)^n N(w; 0, \sigma) \mathrm{d}w$$

$$= \sum_{\substack{i=0 \\ i \text{ even}}}^{n} A_{n,i} \ x^{n-i} \sigma^i - 2\sum_{i=0}^{n} \binom{n}{i} x^i \underbrace{\int_{-x}^{\infty} w^{n-i} N(w; 0, \sigma) \mathrm{d}w}_{J_{-x}(n-i)} \ . \tag{8}$$

We then need to process the expression marked $J$ using IBP, and unlike before we have boundary terms:

$$J_{-x}(p) := \int_{-x}^{\infty} w^p N(w; 0, \sigma) \mathrm{d}w$$

$$= \left[ (-1)^{p-1} \frac{(p-1)!!}{\sqrt{2\pi}} e^{\frac{-x^2}{2\sigma^2}} \sum_{\substack{j=1 \\ j \text{ odd}}}^{p} \frac{x^{p-j} \sigma^j}{(p-j)!!} \right] + \begin{cases} 0, & \text{if p odd} \\ \frac{(p-1)!!}{2} \sigma^p \left[ 1 + \operatorname{Erf}\left( \frac{x}{\sigma\sqrt{2}} \right) \right], & \text{if p even} \ . \end{cases}$$

We can now expand the relevant sum from (8), separating out the Erf parts of $J$ from the sums:

$$-2\sum_{i=0}^{n}\binom{n}{i}x^i J_{-x}(n-i) = -2\sum_{\substack{i=0\\n-i \text{ even}}}^{n}\binom{n}{i}x^i\frac{(n-i-1)!!}{2}\sigma^{n-i}\left[1+\text{Erf}\left(\frac{x}{\sigma\sqrt{2}}\right)\right]$$

$$-2\sum_{i=0}^{n}\binom{n}{i}x^i(-1)^{n-i-1}\frac{(n-i-1)!!}{\sqrt{2\pi}}e^{\frac{-x^2}{2\sigma^2}}\sum_{\substack{j=1\\j \text{ odd}}}^{n-i}\frac{x^{n-i-j}\sigma^j}{(n-i-j)!!}$$

$$= -\left[1+\text{Erf}\left(\frac{x}{\sigma\sqrt{2}}\right)\right]\sum_{\substack{i=0\\i \text{ even}}}^{n}A_{n,n-i}\,x^i\sigma^{n-i}$$

$$+\sqrt{\frac{2}{\pi}}e^{\frac{-x^2}{2\sigma^2}}\sum_{i=0}^{n}(-1)^i A_{n,n-i}\sum_{\substack{j=1\\j \text{ odd}}}^{n-i}\frac{x^{n-j}\sigma^j}{(n-i-j)!!}\,,$$

using the fact that $n$ is even to recondition the sums and simplify the alternating negative sign. We now notice that we can combine all terms of constant $j$ in the second line, rearranging the order of the sums:

$$\sqrt{\frac{2}{\pi}}e^{\frac{-x^2}{2\sigma^2}}\sum_{i=0}^{n}(-1)^i A_{n,n-i}\sum_{\substack{j=1\\j \text{ odd}}}^{n-i}\frac{x^{n-j}\sigma^j}{(n-i-j)!!} = \sqrt{\frac{2}{\pi}}e^{\frac{-x^2}{2\sigma^2}}\sum_{\substack{j=1\\j \text{ odd}}}^{n-1}x^{n-j}\sigma^j\sum_{i=0}^{n-j}(-1)^i\frac{A_{n,n-i}}{(n-i-j)!!}\,.$$

So we may rename index variables $i \to k$ and $j \to i$ and define another constant:

$$B_{n,i} := \sqrt{\frac{2}{\pi}}\sum_{k=0}^{n-i}(-1)^k\frac{A_{n,n-k}}{(n-k-i)!!} = \sqrt{\frac{2}{\pi}}\sum_{k=0}^{n-i}(-1)^k\binom{n}{k}\frac{(n-k-1)!!}{(n-k-i)!!}$$

in order to reunify the sums for even and odd $i$:

$$-2\sum_{i=0}^{n}\binom{n}{i}x^i J_{-x}(n-i) = \sum_{i=0}^{n}x^{n-i}\sigma^i\cdot\begin{cases}-A_{n,n-i}\left[1+\text{Erf}\left(\frac{x}{\sigma\sqrt{2}}\right)\right], & i \text{ even}\\ B_{n,i}\,e^{\frac{-x^2}{2\sigma^2}}, & i \text{ odd}\end{cases}\,.$$

And finally recombine with the first term of (8) to get $F$ itself:

$$F(x|\sigma,n)|_{n \text{ even}} = \sum_{\substack{i=0\\i \text{ even}}}^{n}A_{n,i}\,x^{n-i}\,\sigma^i + \sum_{i=0}^{n}x^{n-i}\sigma^i\cdot\begin{cases}-A_{n,n-i}\left[1+\text{Erf}\left(\frac{x}{\sigma\sqrt{2}}\right)\right], & i \text{ even}\\ B_{n,i}\,e^{\frac{-x^2}{2\sigma^2}}, & i \text{ odd}\end{cases}$$

$$= \sum_{i=0}^{n}x^{n-i}\sigma^i\cdot\begin{cases}A_{n,i}-A_{n,n-i}\left[1+\text{Erf}\left(\frac{x}{\sigma\sqrt{2}}\right)\right], & i \text{ even}\\ B_{n,i}\,e^{\frac{-x^2}{2\sigma^2}}, & i \text{ odd}\end{cases}\,. \quad (9)$$

In particular, for the $n = 2$ case which might have utility for modeling drag amidst turbulence, we have

$$F(x|\sigma,2) = -\left(\sigma^2+x^2\right)\text{Erf}\left(\frac{x}{\sigma\sqrt{2}}\right) - \sqrt{\frac{2}{\pi}}x\sigma e^{\frac{-x^2}{2\sigma^2}}\,,$$

$$\implies G(x|\sigma,2) = \sqrt{\mu_{4,x,\sigma}+3F^2(x|\sigma,2)}$$

$$= \sqrt{x^4+6x^2\sigma^2+3\sigma^4+3\left[\left(\sigma^2+x^2\right)\text{Erf}\left(\frac{x}{\sigma\sqrt{2}}\right)+\sqrt{\frac{2}{\pi}}x\sigma e^{\frac{-x^2}{2\sigma^2}}\right]^2}\,.$$

## S2. ADDITIONAL FIGURES

Here we have provided some additional 3-dimensional figures elaborating on the 2-dimensional ones from the main text.

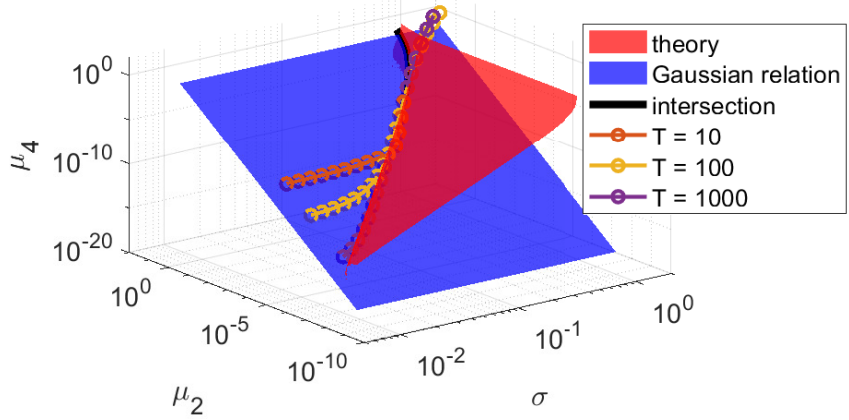

FIG. S1: **Small-noise convergence.** Three dimensional version of Fig. 1 (top panel) from the main paper; at small noise $\sigma$, solutions converge over increasing simulated time $T$ to the intersection of the Gaussian condition and Eq. (11), our theorized surface from the main paper relating the equilibrium's second and fourth moments to the inherent noise $\sigma$: $0 = 15\sigma^6 + 6\sigma^2(6\sigma^2 - 1)\mu_2 + (9\sigma^2 - 2)\mu_4$.

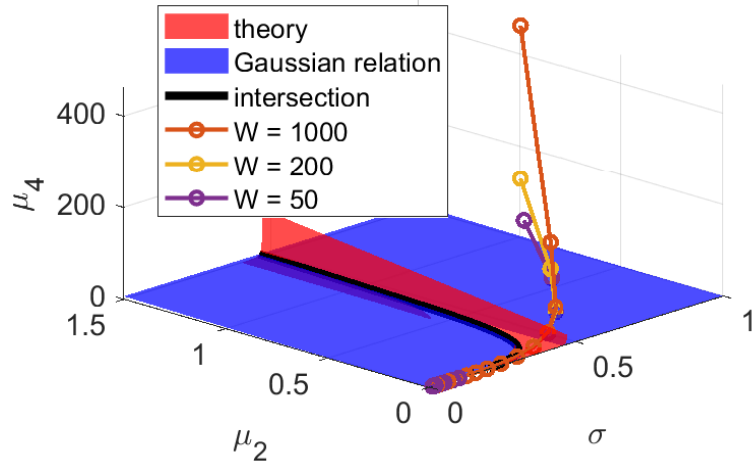

FIG. S2: **Large-noise divergence.** Three dimensional version of Fig. 1 (bottom panel) from main paper relating noise amplitude $\sigma$ to the second and fourth moments of the equilibrium ($\mu_2$ and $\mu_4$, respectively), but on a linear scale. We can see the clear suggestion of divergence for (at least) $\mu_4$; as our domain captures more standard deviations $W$ of the solution, the measured $\mu_4$ grows without bound. We propose that noise values $\sigma$ beyond the asymptote "curtain" $\sigma^* = \sqrt{2}/3 \approx 0.47$ must have divergent second and fourth moments.

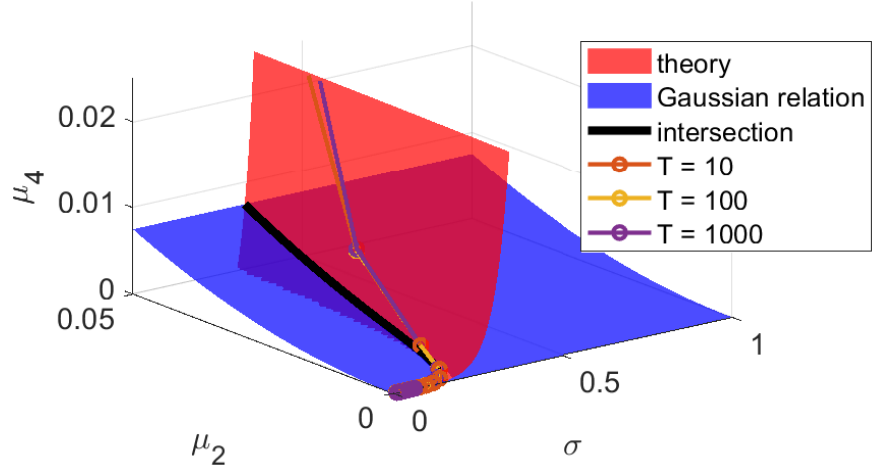

FIG. S3: **Gaussian validity boundary.** Three dimensional, linear-scale zoom of the region where the Gaussian assumption fails to hold. Our solutions fall off the intersection line due to their non-Gaussian nature.

## S3. GAUSSIAN/NON-GAUSSIAN TRANSITION

In the main text we argue that the small-noise limit of the stochastic-cubic-attractor system leads to a Gaussian equilibrium. We can see this transition clearly in Fig. S4.

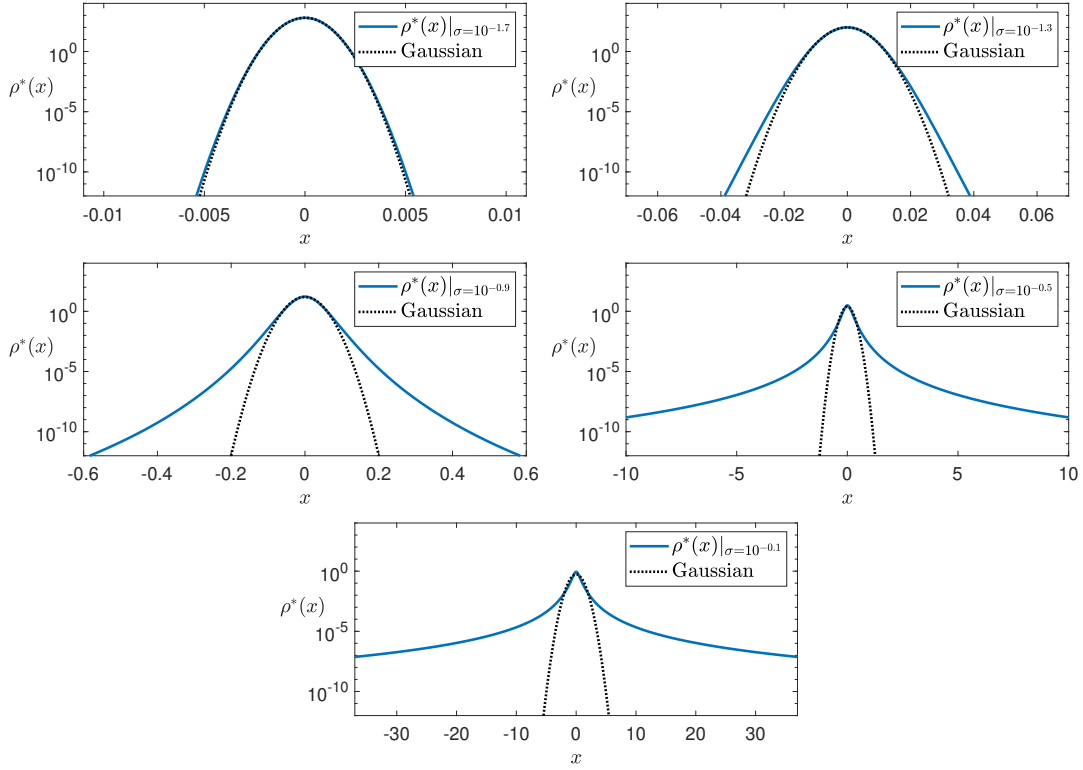

FIG. S4: **Equilibrium shape transition.** Comparison of equilibrium distributions for different noise values to Gaussian distributions with the same standard deviation. **Top Left:** For small $\sigma$, equilibria exhibit the signature parabolic shape (on semi-log axes) indicating a near-Gaussian distribution. **Other panels:** For larger $\sigma$, equilibria exhibit increasingly "fat tails" differentiating them from Gaussians.



\end{document}